\documentstyle[aps,prl,preprint,epsfig]{revtex}
\tightenlines
\begin{document}
\title{{\bf Lyapunov Exponent, Generalized Entropies and Fractal Dimensions Of
 Hot Drops}}
\author{C. O. Dorso$^{1,2}$ and A. Bonasera$^1$}
\address{{\it $^1$}
Laboratorio Nazionale del Sud - Istituto Nazionale di Fisica Nucleare,\\
via S. Sofia 44, I-95123 Catania, Italy }
\address{{\it $^2$Departamento de Fisica, Facultad de Ciencias Exactas y}\\
Naturales Universidad de Buenos Aires\\
Pabellon I, Ciudad Universitaria, Nu\~nez\\
1428 Buenos Aires, Argentina}
\maketitle

\begin{abstract}
We calculate the maximal Lyapunov exponent, the generalized entropies, the
asymptotic distance between nearby trajectories and the fractal dimensions
for a finite two dimensional system at different initial excitation
energies. We show that these quantities have a maximum at about the
same excitation energy. The presence of this maximum indicates the
transition from a chaotic regime to a more regular one. In the chaotic
regime the system is composed mainly of a liquid drop while the regular
one corresponds to almost freely flowing particles and small clusters. 
At the transitional
excitation energy the fractal dimensions are similar to those estimated from
the Fisher model for a liquid gas phase transition at the critical point.
\end{abstract}

 \newpage
\noindent

Infinite systems composed of particles interacting with an attractive plus, a
shorter range, repulsive force have an Equation Of State (EOS) resembling a
Van der Waals one \cite{pp3}, which exhibits phase transitions from solid to
liquid and/or to gas. The features of the EOS of such a system are quite
independent on the specific form of the two body potential, i.e. a sum of
Yukawa's or Lennard-Jones potential etc.
A problem arises when the system is constituted
of a finite number of particles $N$ and it is not confined in a box. In such
a limit, it is not strictly correct to define a critical point, on the other 
hand it becomes very interesting to analyze how the
system behaves as a function of its excitation energy. Intuitively we expect
that at low excitation energies a transition, from solid-like state to
liquid-like state, for a finite system should be very similar to the
infinite case limit. This is so because at these low energies
the attractive part of the potential is dominant and the system
remains confined in a given, self sub-stained, volume. Thus it has sufficient
time to develop correlations that are characteristic of such a
phase transition\cite{pp10}. In fact in this regime the Caloric Curve (i.e.
the temperature $T$ of the system as a function of the excitation energy
 $E^{*}$) displays
the standard ''rise-plateau-rise'' pattern around the solid-like to
liquid-like transition (i.e. the solid branch, the coexistence region and
the liquid branch)\cite {labastie}. At higher excitation energies 
the system is unable
to remain confined and undergoes a fragmentation process. This kind of
process is characterized by the appearance of a new degree of freedom, the
one associated with the collective expansion. In this case,  it has been found
that many features of a thermodynamical liquid-gas transition are 
reproduced even if
the system has mass as low as $A=100$ \cite{pp9}. These features are mainly
deduced from the analysis of asymptotic mass distributions and in particular
one finds a power law in the mass yield for a given initial excitation
energy. There have been also estimates of the critical exponents from 
data in nucleus-nucleus and cluster-cluster 
collisions  \cite{eos,fari}. On the other hand the
corresponding Caloric Curve does not show the usual increase in the
temperature of the so called "vapor branch" with the increase of the 
excitation energy, but
instead a plateau is reached as soon as the systems enters the fragmentation 
regime%
\cite{strador}.  The Maximal Lyapunov Exponent (MLE) has been
studied in Classical Molecular Dynamics (CMD) for a 3 dimensional system
composed of 100 particles and for different initial excitation energies. In 
\cite{pp2} a maximum in the MLE was found for an initial excitation energy
where a power law in the mass yield, intermittency signal, largest variance
in the size of the biggest fragment\cite{ppd,pp9,cam86}
 are also obtained. It is the
purpose of this letter to strengthen and better characterize this result by
analyzing the behavior of other important indicators of chaoticity, i.e.
the asymptotic distance between trajectories \cite{ther,bar}, the
Generalized Renyi's Entropies (GRE) and the fractal dimension \cite{pp1}. We
will solve the classical equation of motion (CEOM) for a system composed by 
100 particles interacting through a 6-12 Lennard Jones potential in d=2
dimensions. Details on the method of solution of the CEOM and the
preparation of the initial state are given in \cite{sd1}.

In order to calculate the MLE, we generate at time t=0, for each trajectory,
a second one where we change the momenta of the particles by a small amount $%
d_0$ in momentum space. Following \cite{pp2} we define a distance between
trajectories $d(t)$~ as: 
\begin{eqnarray}
d(t)=~\left( \frac 1N\sum_{i=1}^N[a({\bf r}_1(t)-{\bf r}_2(t))^2+b({\bf p}%
_1(t))-{\bf p}_2(t))^2]_i\right) ^{1/2}~~~\ ,  \label{eq1}
\end{eqnarray}
where ${\bf r},{\bf p}$ refer to the positions and momenta of $N$~ particles
at time $t$. Indices $^{\prime }1^{\prime }$ and $^{\prime }2^{\prime }$
refer to the two trajectories differing by $d_0$~ at $t=0$~. $a,b$ are two
arbitrary parameters which express the fact that the LE are independent of
the particular metrics in the phase space\cite{note}.  For the purpose
of this paper we will fix $a=0,b=1/m$ where m is the mass of the particles,
i.e. distances in velocity v-space only. If we calculate numerically the
time evolution of d(t) solving the CEOM, we observe an exponential increase
followed by saturation in v-space \cite{pp2,ther,bar}. The exponential
increase of d(t) is associated to the MLE ${\hat \lambda }$ and it implies
the following relation $\frac{d [d(t)]}{dt}=\hat \lambda d(t)$.
But this rapid increase cannot last forever because the available v-phase
space is limited, giving rise to a saturation of the inter trajectory 
distance in v-space.
In order to describe this saturation, we can consider the previous relation 
as a first order
term in an expansion in d(t), going to second order we get\cite{ther}: 
\begin{eqnarray}
\frac{d [d(t)]}{dt}=\hat \lambda d(t)-\alpha d^2(t)+..~~~~\ ,
\label{eq2}
\end{eqnarray}
where $\alpha $~ is a constant greater than zero for fully developed
chaos. Eq. (\ref{eq2}) can be easily solved, giving: 
\begin{eqnarray}
d(t) &=&\frac{d_\infty d_0}{d_0+d_\infty e^{-\hat \lambda t}}  \label{eq4} \\
&&  \nonumber \\
\hat \lambda  &=&\alpha d_\infty ~~~~~\ ,
\end{eqnarray}
where $d_0=d(t=0)$ and $d_\infty =d(t=\infty )$. Thus, eqs.(3,4) tell us
that to characterize the entire time evolution of d(t), we need three
quantities, $\lambda $, the asymptotic distance between trajectories $%
d_\infty $ and $\alpha $, but only two quantities are independent because of
eq.(4). In particular since $\alpha $ is a constant we find that the LE are
proportional to $d_\infty $. In ref. \cite{ther} this relation was supported
from numerical simulation in Hamiltonian systems, similarly in ref.\cite{bar}
for maps. 

The MLE is proportional to the distance in v-space
which provides a measure of the fluctuations.  
For instance for an infinite system in equilibrium 
as $t\rightarrow \infty $,
 the momenta of
particles in event '1' are uncorrelated to
those of event '2', and it is very easy to show that in such cases the $%
d_\infty $ is proportional to the variance in v-space. This is a very useful
result which allows us to estimate the LE using the final distributions
obtained either from the data or from the theory such as the thermodynamics.
For example, for a classical Boltzmann gas the variance of the velocity
distribution $\sigma $ is given by\cite{pp3}~: 
\begin{eqnarray}
d_\infty \propto \sigma =\left( {\frac{3T}m}\right) ^{1/2}~~~~~\ ,
\label{eq6}
\end{eqnarray}
where $T$ is the temperature of the gas measured in units of energy.
 For the infinite system the MLE is
then an increasing function of the temperature of the system \cite{ther,po1}%
. On the other hand, in the case of a free expansion of a finite system
(collective motion), $d_\infty =d_0$ holds, i.e. $\hat \lambda =0$.

In figure (1) we plot the MLE and the $d_\infty $ versus $E^{*}$ as obtained
in our CMD simulations. 
 The qualitative features are the same as those obtained in ref. 
\cite{pp2,ther}.  
Both quantities plotted display a maximum even though at slightly
different $\epsilon$. The decrease of the $d_\infty $ for large $\epsilon$, 
suggests that the particles having an initial kinetic energy
larger than the binding energy escape quickly from the system without
interacting. In fact if we compare this figure with the caloric curve
displayed in ref.\cite{strador} we will notice that $d_\infty $ attains
its maximum when the caloric curve reaches the plateau, which signals the
state at which the dynamics of the system begins to be dominated by the
collective radial flow. Similar considerations apply to the MLE. This 
supports also the idea of a limiting temperature that a finite system
can sustain \cite{wada}.

The maximum in the MLE signals a transition from a chaotic to a more ordered
motion i.e. a motion in which the expansion collective mode is more an more
important. For a finite system the main effect of collective motion is to
suppress inter-particle collisions, in fact the higher the initial energy the
faster the systems breaks and the smaller the final fragments are. Such a
behavior resembles the one that has already been observed in \cite{po2} in a
liquid to solid transition for the correlated cell model when changing the
density. Notice indeed the similarity of the two cases. Small $\epsilon$ in our
case corresponds to small $\rho $ in \cite{po2} i.e. the chaotic motion
occurs in the liquid. At high $\epsilon$, we obtain a more ordered motion
because of the collective expansion, while in \cite{po2} at high $\rho $ the
system becomes a solid displaying regular trajectories which remains trapped
within the volume determined by the neighboring particles. 

If our simulations are followed for a long time, stable fragments will finally
be formed. From the mass distributions we can estimate the GRE as follows.
Define the probability of finding a fragment of mass i as the number of
fragments $M(i,\delta )$, where $\delta $ is the mass resolution, divided
the total number of fragments produced for a given event at that $\epsilon$.
Thus 
\begin{eqnarray}
p(i,\delta )={\frac{M(i,\delta )}{\sum M(i,\delta )}}~~~~~\ .  \label{eq7}
\end{eqnarray}
The GRE are \cite{pp1}: 
\begin{eqnarray}
S_q(\delta )={\frac 1{1-q}}log({\sum_i<p_i>}^q)~~~~~\ .  \label{eq8}
\end{eqnarray}
Where $<>$ denote the average over an ensemble and we take $q$ as an integer
number. It is important to stress that the minimum mass resolution possible
for finite systems is clearly $\delta =1$.

In figure (2a) we plot $S_q(\delta =1)$ vs. $\epsilon$ 
for $q=0-6$: a clear peak
is observed.  This peak is precisely in the region in which the MLE also 
shows a peak.  Note that in the calculation we restricted the sum to those
particles having mass larger than 2. If we keep smaller masses the peak
remains even though it broadens. 

 From the knowledge of the GRE we can define the generalized dimensions (GD)
as: 
\begin{eqnarray}
D_q={\frac{lim}{\delta \rightarrow 0}}{\frac{S_q(\delta )}{log\delta }}%
~~~~~\ ,  \label{eq9}
\end{eqnarray}
i.e. we study the way in which $S_q(\delta )$ scales with $\delta$\cite{note1}.
 In 
figure (2b) we show the fractal dimension ${D_q}$ vs. $\epsilon$. It is once 
again immediate to see a peak in ${D_q}$ ($q \ge 1$)
in the same region in which
MLE and ${S_q}$ displayed a peak. Finally in  
figure (3) we plot the ${D_q}$ vs. $q$ for various excitation energies. For
illustration we discuss some limiting cases. For instance if the mass
distribution is uniform, we easily get $D_q=1$ for all q. This is a trivial
case which tells us that the entire space is uniformly covered and the $D_q$
are equal to the topological dimension 1. Another limiting case is when all
the particles are concentrated in one bin (say mass 1) and zero otherwise.
This gives $D_q=0$ which is the dimension of the space occupied i.e. the
dimension of a point.  It is also interesting to note that if the $p_i$s are
different from $0$ for M contiguous bins only, then 
$D_0=1$ ,i.e.  the Hausdorff dimension of a segment.
A more interesting case is when the mass distribution
is given by a power law. We can write such a mass distribution as $%
y(x)\propto x^{-\tau }$ where $x\epsilon [\epsilon ,1]$, $x=i/N$ and $%
\epsilon $ is a small quantity related to the smallest possible mass that we
can have. Following \cite{pp1}, taking the limits $\delta \rightarrow
\epsilon \rightarrow 0$ gives, for $\tau <1$:

$D_q$=1  if $q<1/\tau$;   $D_q$=$q(1-\tau)/(q-1)$  if $q\ge 1/\tau$

For instance, $\tau=0.5$ gives the GD as for the logistic map at r=4 \cite
{pp1}, with $D_q$ continuous but its first derivative has a discontinuity at $%
q=2$ and this behavior is referred to as a first order phase transition. 
%
%
As we noticed before in our case we get a power law distribution for the
excitation energy where the MLE and the GRE have a maximum.  Since the power
that we get is larger than 2 it is interesting to see what the behavior
of the $D_q$ is in such a case which corresponds to a second order
phase transition (at least in the infinite case limit). 
First we have simulated numerically a power law yield and the
result is plotted in fig.(3) (full squares), with $A=100$ and $\tau =2.1$. The
 CMD results for $\epsilon=-0.75$(circles),
where a power law in the mass distribution is obtained,
 are in good agreement with
the simulation.  We notice that the $D_q$ show no discontinuities at variance
with the cases where $\tau \le 1$.  In order to test if this is a finite
size effect we have simulated fragmentation in a simple percolation model
whose properties are well studied.  We find a similar behavior to the one
discussed above at the critical percolation point and for very large sizes,
more details will be discussed elsewhere.  
The $D_q$ are also
plotted for the cases when $d_\infty$ has a maximum (open circles) and at
high $\epsilon$ (open squares). 
The functional dependence of the $D_q$ with q suggests
a multi fractal character of the probability distributions.

 From these analysis the behavior of excited finite systems is greatly
clarified. If we start with a cold (solid) drop of matter and
 begin to heat it up the Lyapunov exponent increases as well as the $%
d_\infty $. The first one is proportional to the rate at which the system
explores phase space and the second to the available phase space. This
trend changes as the rate of evaporation increases i.e when radial flow, the
extra degree of freedom characteristic of the evolution of highly excited 
finite systems, starts to play a
major role in the evolution. A maximum of both quantities is reached when
the system approaches the critical region, i.e. when fluctuations are maximal
and the final spectra contain both liquid-like and vapor like components.
For even higher excitation energies both values decrease as a result of
the fast fragmentation process and the transfer of chaotic (thermal) energy
into ordered (radial flux) energy. A peak is also observed in the
generalized Renyi entropies and in the generalized Fractal Dimensions ${D_q}$. 
The $D_q$ at the 'critical' point for which a power law with $\tau \ge
 2$ is obtained is a smooth decreasing function of $q$, at variance with
the cases where $\tau \le 1$.

\newpage 

\begin{figure}[tbp]
\begin{center}
\mbox{{ \epsfysize=12 truecm \epsfbox{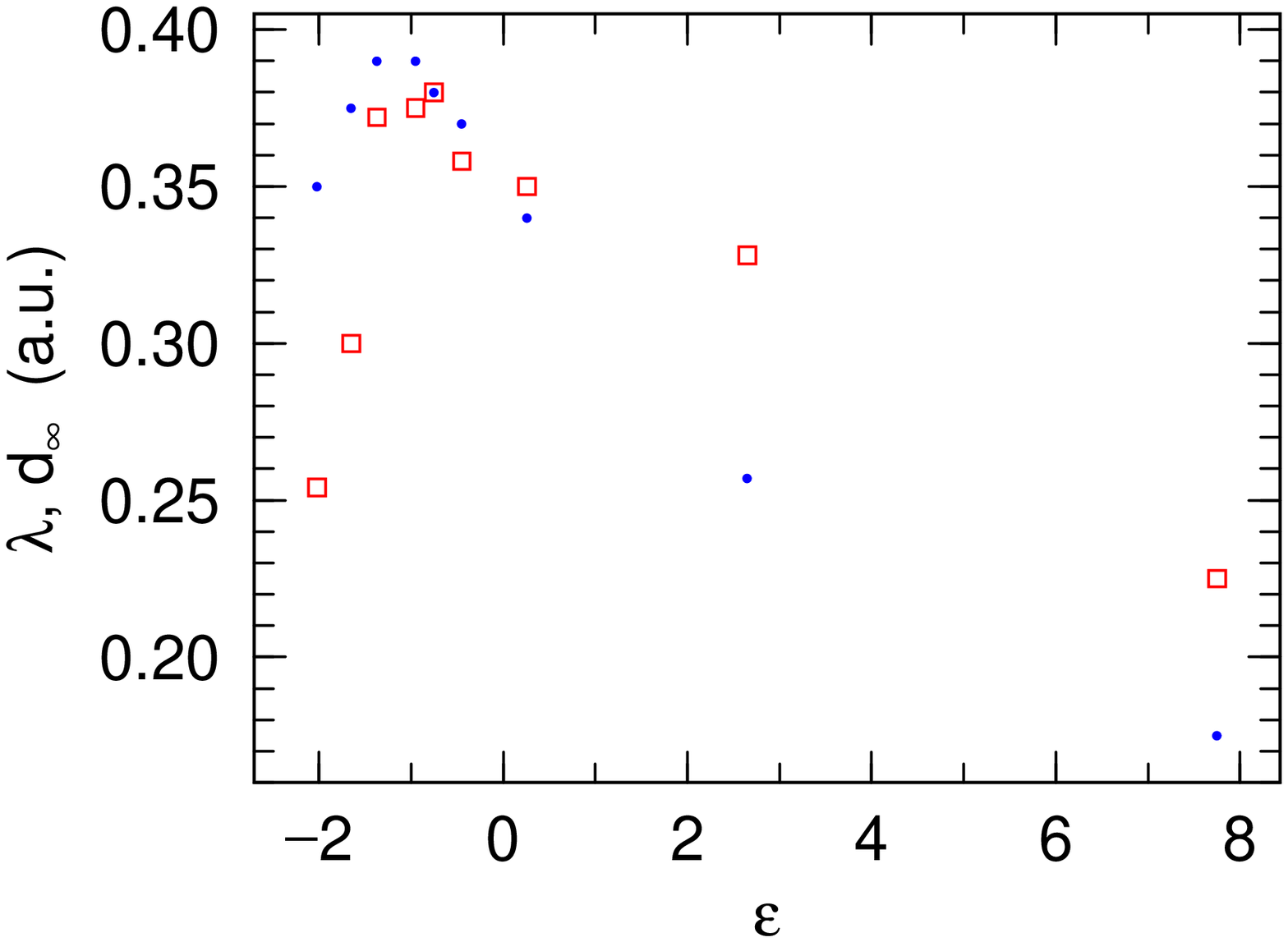}}}
\vskip 1.5cm
\label{f1} \noindent
\caption{ MLE (squares) and $d_{\infty}/4$ (dots) as a function of energy
(in natural units $\epsilon$ [14]) for two dimensional L.J. drops with N=100
particles.}
\end{center}
\end{figure}
\begin{figure}[tbp]
\begin{center}
\mbox{{ \epsfysize=12 truecm \epsfbox{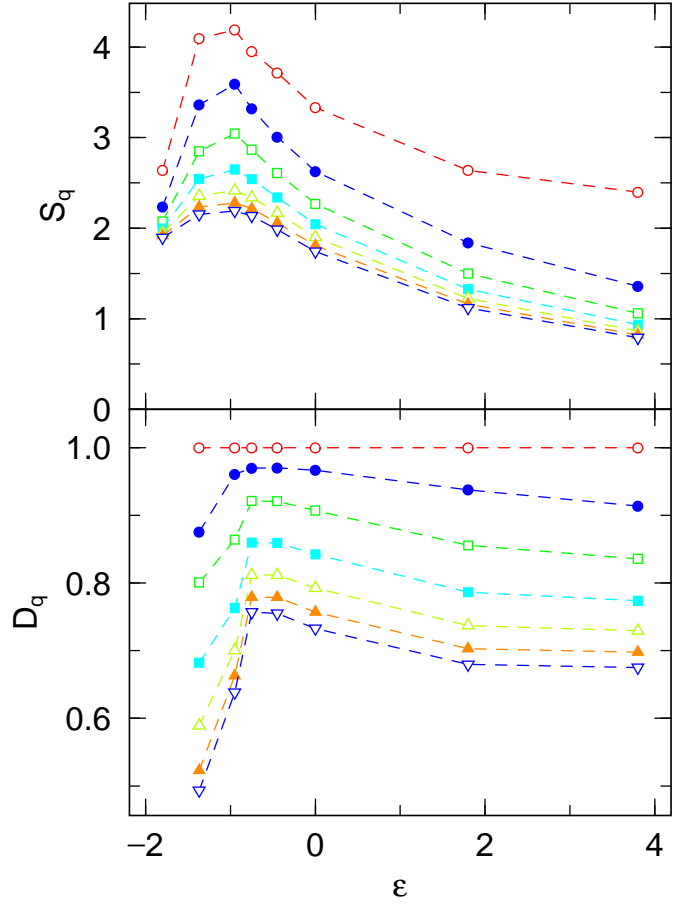}}}
\vskip 1.5cm
\label{f2} \noindent
\caption{ Generalized entropies (top) and fractal dimensions (bottom)
 vs. energy
calculated from the asymptotic mass spectra
for the same system as Fig.1. Open circles q=0  , full circles q=1,
 open squares q=2 , full squares q=3 , open triangles q=4, full triangles q=5,
 down pointing triangles q=6}
\end{center}
\end{figure}
\begin{figure}[tbp]
\begin{center}
\mbox{{ \epsfysize=12 truecm \epsfbox{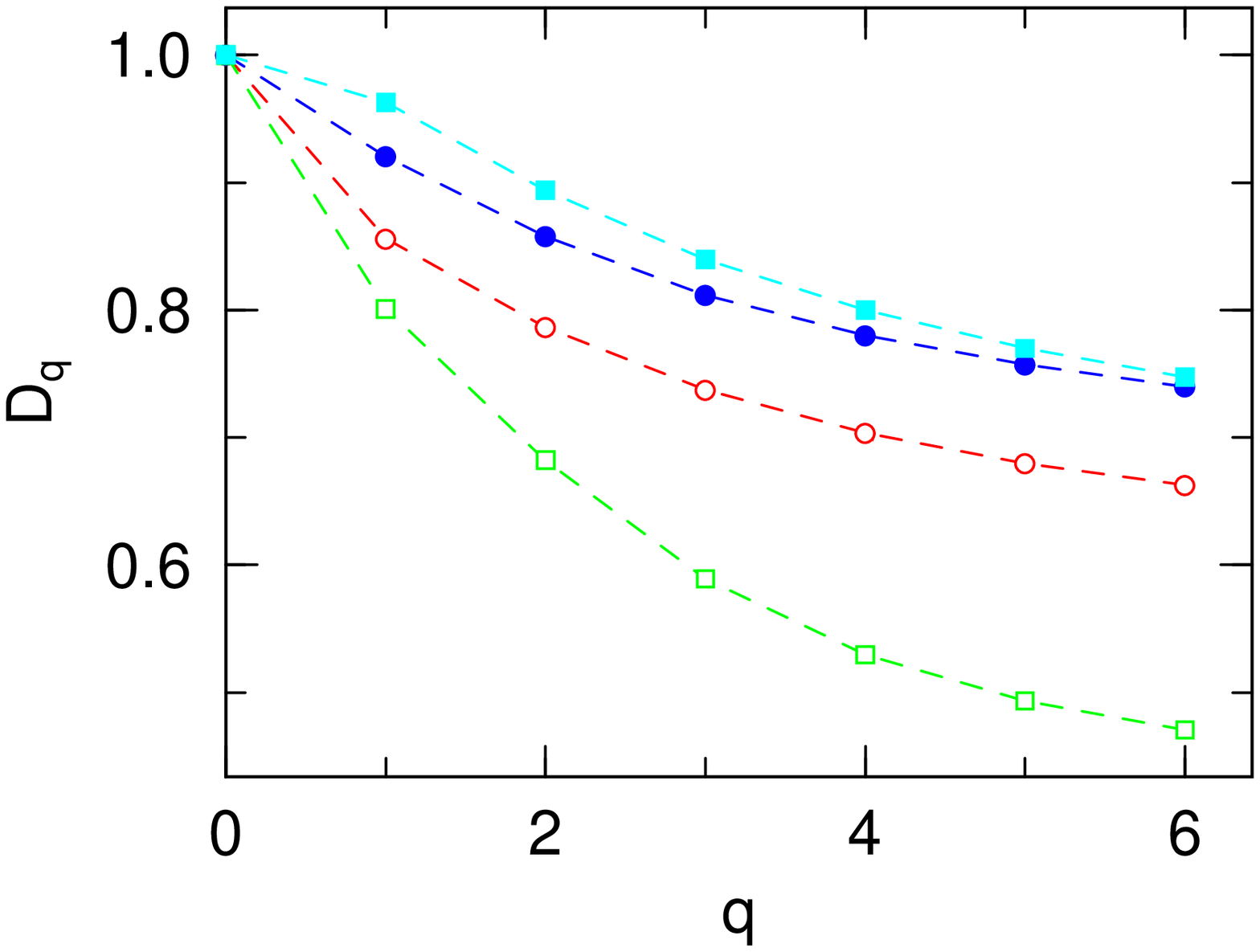}}}
\vskip 1.5cm
\label{f3b} \noindent
\caption{Generalized fractal dimensions at $\epsilon=-1.37 $ (open circles),
 $%
-0.75 $ (full circles), and $2.25 $ (open squares). The full 
squares are
obtained assuming a power law mass distribution.}
\end{center}
\end{figure}



\begin{thebibliography}{99}
\bibitem{note}  We remark that if the collective effects are important, then 
the calculations of the MLE become sensitive to the time when the second
trajectory is generated. Here, we are always referring to the MLE as
obtained by generating a second trajectory at $t=0$[8].
\bibitem{note1} Notice that even though the minimum possible mass
is $\delta =1$, when plotting $S_q(\delta)$ vs. $\delta$ in log-log
 scale we can extrapolate to $\delta \rightarrow 0$. We would like
to stress that the range where the fit is performed , see eq(8), is
limited due to the finiteness of the system.
\end{thebibliography}
\end{document}